\documentclass[RNAAS]{aastex7}

\usepackage{amsmath}
\RequirePackage[fixed]{fontawesome5}
\newcommand{\github}[1]{%
   \href{#1}{\faGithub}%
}
\shorttitle{PowerSpectR}
\shortauthors{de Souza}
\graphicspath{{./}{figures/}}

\begin{document}

\title{PowerSpectR: An R Package for Radial Power Spectrum Estimation}

\correspondingauthor{Rafael S. de Souza}

\author[0000-0001-7207-4584]{Rafael S. de Souza}
\email[show]{rd23aag@herts.ac.uk}
\affiliation{Centre for Astrophysics Research, University of Hertfordshire, College Lane, Hatfield, AL10 9AB, UK}
\affiliation{Instituto de Fisica, Universidade Federal do Rio Grande do Sul, Porto Alegre, RS 90040-060, Brazil}
\affiliation{Department of Physics \& Astronomy, University of North Carolina at Chapel Hill, NC 27599-3255, USA}

\begin{abstract}

I present here \texttt{PowerSpectR}, an \textsf{R} package for computing and visualizing median-based radial Fourier power spectra from imaging data. Power spectra provide a representation of spatial structure by decomposing contributions across spatial scales, and the resulting slopes can serve as compact, low-dimensional summaries of morphological complexity across images.
\texttt{PowerSpectR} provides a workflow for estimating these slopes, combining edge-effect mitigation through  Hann windowing \citep{blackman1958measurement}, Fourier-domain analysis, and radial binning with azimuthal median statistics. The use of median aggregation helps to reduce sensitivity to bright compact sources, masking artifacts, and other localized features that can bias standard estimators. \texttt{PowerSpectR} is released under the MIT license at \href{https://github.com/RafaelSdeSouza/PowerSpectR}{this repository \faGithub},  and the archived software release is available in Zenodo \citep{PowerSpectR_v010}.
\end{abstract}


\section{Introduction}

Two-dimensional power spectra are widely used to quantify the spatial structure of astronomical images, where the fitted slope encodes how power is distributed across spatial scales \citep[e.g.,][]{Elmegreen2004, Lazarian2000, Burkhart2013}. By mapping an image into the frequency domain, spatial variations at different scales are explicitly separated, and the resulting distribution of power provides a translation-invariant representation of morphology, and—through radial averaging—an isotropic summary of spatial structure.

Related applications of power-spectrum-based characterization in astronomy include the detection of hostless transients \citep{Pessi2024}, the characterization of young stellar object environments \citep{Kuhn2021}, and the identification of multiply imaged, gravitationally lensed quasars \citep{Stern_2021}. Despite the broad applicability of power-spectrum-based descriptors across astrophysical data, this interpretation is not restricted to astronomy: similar scale-dependent behavior arises in natural images and textured fields, where power spectra capture generic properties of spatial organization \citep{VANDERSCHAAF1996}. 

Motivated by prior work and by similar \textsf{Python} implementations \citep[\texttt{TurbuStat};][]{Koch2016,Koch2019}, I introduce \texttt{PowerSpectR}, a lightweight \textsf{R} package for median-based radial power-spectrum estimation.

\section{Methodology}

Given an image \(I(x, y)\) with dimensions \(h \times w\), the package first subtracts the image mean and applies a two-dimensional Hann window \(W(x, y)\) to suppress edge effects. The power spectrum is then
\begin{equation}
P(k_x, k_y) \propto \left| \mathcal{F}\left\{ W(x,y)\,[I(x,y) - \bar{I}] \right\} \right|^2,
\end{equation}
where \(\mathcal{F}\{\cdot\}\) denotes the discrete Fourier transform, and the proportionality constant depends on the normalization convention of the discrete Fourier transform and does not affect the estimated spectral slope. 
After centering the Fourier domain (FFT shift), the radial coordinate is defined as
\begin{equation}
k = \sqrt{k_x^2 + k_y^2}.
\end{equation}
The power spectrum is then binned and summarized using the azimuthal median,
\begin{equation}
\tilde{P}(k_j) = \mathrm{median}\!\left\{ P(k_x, k_y)\,:\, k \in [k_j^{\mathrm{min}}, k_j^{\mathrm{max}}] \right\}.
\end{equation}
Assuming a power-law scaling of the form $\tilde{P}(k) \propto k^{\beta}$, the fitted spectral slope $\beta$ is obtained from a linear model in log--log space,
\begin{equation}
\log \tilde{P}(k) = \alpha + \beta \log k + \varepsilon,
\end{equation}
where \(\varepsilon\) captures stochastic fluctuations around the trend.

\section{Analysis}
I illustrate the method using two Hubble Space Telescope image cutouts representing distinct morphological classes. For each image, I compute the median radial power spectrum and estimate the spectral slope \(\beta\).

\begin{figure}[h]
\centering
\includegraphics[width=0.7\linewidth]{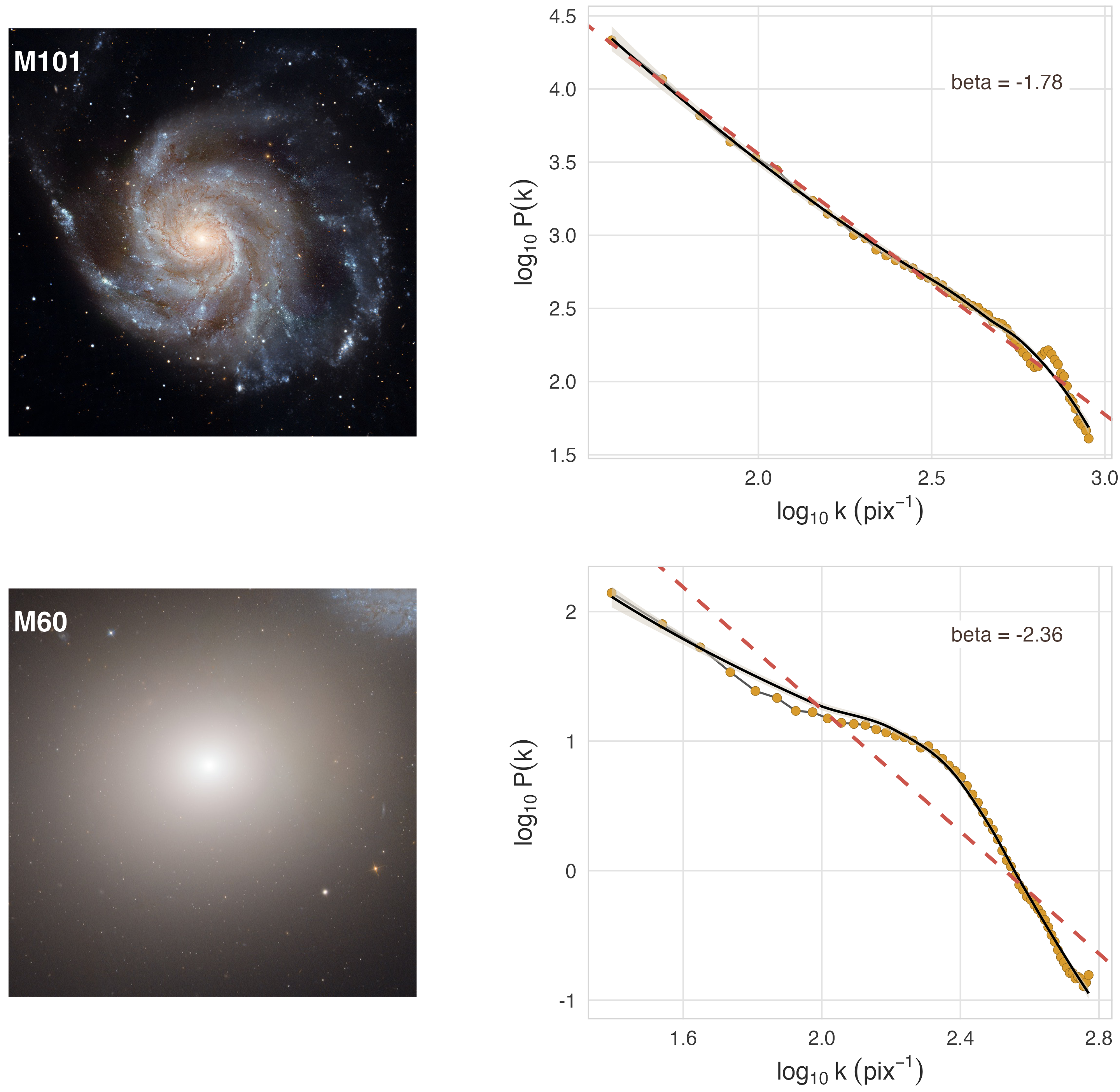}
\caption{Comparison of Hubble Space Telescope image cutouts of M101 (spiral) and M60 (elliptical). Left: images. Right: median radial power spectra computed with \texttt{PowerSpectR}. 
\textit{Credit:} M60—NASA, ESA, and the Hubble Heritage (STScI/AURA)-ESA/Hubble Collaboration; 
M101—NASA, ESA, K. Kuntz (JHU), F. Bresolin (University of Hawaii), J. Trauger (JPL), 
J. Mould (NOAO), Y.-H. Chu (University of Illinois, Urbana), and STScI.
\label{fig:hubble-comparison}}
\end{figure}

Figure~\ref{fig:hubble-comparison} shows the corresponding results for M101 and M60. The fitted slopes are \(\beta \approx -1.78\) and \(\beta \approx -2.36\), respectively, reflecting the different spatial structure of the two systems.

The spiral galaxy exhibits a flatter spectrum, indicating relatively enhanced power at small spatial scales, whereas the elliptical galaxy yields a steeper slope consistent with a smoother light distribution. This illustrates how the spectral slope acts as a compact descriptor of morphological complexity on spatial scales.

\section{Conclusions}

\texttt{PowerSpectR} provides a compact and robust implementation of median-based radial power-spectrum estimation for imaging data. By adopting azimuthal median statistics, the method has less  sensitivity to bright compact sources, masking artifacts, and other localized features, yielding stable estimates of spectral slopes.

The package offers an end-to-end workflow—from edge-effect mitigation via Hann windowing to slope estimation in log--log space—facilitating reproducible extraction of scale-dependent structural information. In this context, the spectral slope can be interpreted as a low-dimensional descriptor of morphology, enabling efficient feature mapping across large image datasets. This is particularly relevant for current and upcoming wide-field surveys,  such as the Vera C. Rubin Observatory Legacy Survey of Space and Time \citep{Ivezic2019}, Euclid \citep{EuclidCollaboration2022}, the Chinese Space Station Survey Telescope  \citep{2026SCPMA},  and the Nancy Grace Roman Space Telescope \citep{Akeson2019}, which will produce imaging data at unprecedented scale and depth.

The framework is general and can be applied to astronomical images, emission-line maps, and slices from integral-field spectroscopy, as well as to other classes of spatial data where scale-dependent structure is of interest.

\begin{acknowledgments}
R.S.S. acknowledges support from the Conselho Nacional de Desenvolvimento Cientifico e Tecnologico (CNPq, Brazil, grants 446508/2024-1 and 315026/2025-1). The author thanks the open-source \textsf{R} community for the tools that made this package possible.
\end{acknowledgments}

\software{R \citep{RCoreTeam2025}, imager \citep{Barthelme2025}, ggplot2 \citep{Wickham2016}}

\bibliography{ref}{}
\bibliographystyle{aasjournal}

\end{document}